\documentclass[aps,prb,twocolumn,superscriptaddress]{revtex4}
\usepackage{graphicx}
\usepackage{color}

\usepackage{amssymb}

\begin{document}
\title{Signature of topological crystalline insulating behavior in new B$_2X_2$Zn ($X$=Ir, Rh, Co) compound from first-principles Computation}

\author{J.~Howard} 
\affiliation{Department of Physics, Seton Hall University, South Orange, New Jersey 07079, USA}
\author{A.~Rodriguez} 
\affiliation{Department of Physics, Seton Hall University, South Orange, New Jersey 07079, USA}
\author{N.~Haldolaarachchige} 
\affiliation{Department of Physical Science, Bergen Community College, Paramus, New Jersey 07652, USA}
\author{K.~Hettiarachchilage} 
\affiliation{Department of Physics and Astronomy, College of Staten Island, Staten Island, New York 10314, USA}

\date{\today}

\begin{abstract}
Recent attempts at topological materials have revealed a large class of materials that show gapless surface states protected by time-reversal symmetry and crystal symmetries. Among them, topological insulating states protected by crystal symmetries, rather than time-reversal symmetry are classified as topological crystalline insulators. We computationally predict the signature of new three-dimensional topological crystalline insulating compounds of space group 139(I/4mmm).  After conducting a full volume optimization process by allowing to rearrange of atomic positions and lattice parameters, the first principles calculation with a generalized gradient approximation is utilized to identify multiple Dirac-type crossings around X and P symmetric points near Fermi energy. Importantly the band inversion at point P is recognized. Further, We investigate the compound for topological crystalline insulating behavior and identify metallic surface states on high-symmetry crystal surfaces with the projection to the plane (001). Additionally, we performed formation energy, elastic properties, and phonon modes calculations to verify the structural, mechanical, and dynamical stability of the compounds. Therefore, we suggest the compounds for further investigation and experimental realization.
\end{abstract}
\maketitle

\section{Introduction}
Following the discovery of a two-dimensional topological insulating phase in graphene with a significant effect of spin-orbit coupling (SOC)~\cite{Kane, Kane2} rigorous efforts have been committed to exploring new topological materials. Among them, the identification of topological insulators (TIs) and topological semimetals (TSMs) with unusual physical properties and potential applications lead to enthralling discoveries both theoretically and experimentally. TIs presence of a bulk insulating gap and gap-less conducting surface states that form a Dirac point ~\cite{Hasan, Qi} while topological semimetals (Dirac semimetals, Weyl semimetals, and nodal line semimetals) exhibit gapless bulk states and band crossings in momentum space which are inverted beyond the crossing point (or line)~\cite{Yang1, Xu1, Zyuzin} protected by topology and symmetry.

The gapless surface states across the band gap in both two- and three-dimensional TIs characterized by a $Z_2$ topological invariant, are topologically protected by time-reversal symmetry~\cite{Hasan, Qi, Mong, Teo}. The binary alloys $Bi_2Se_3$, $Bi_2Te_3$, and $Sb_2Te_3$ that hold topological surface states are identified as strong TIs~\cite{Xia, Chen, Hsieh1}. Topological crystalline insulators (TCIs) originated as a counterpart of topological insulators without the significant effect of SOC. The crystal symmetries in TCIs play a vital role in the topology of the bands showing gapless surface states across the insulating ~\cite{Teo, Fu, Hsieh, Weng, Fang}. TCI phase was first theoretically predicted and experimentally realized by angle-resolved photoemission spectroscopy in SnTe, $Pb_{1-x}$Sn$_x$Se, and $Pb_{1-x}$Sn$_x$Te~\cite{Hsieh}. An integer
topological invariant known as the mirror Chern number arising from the mirror symmetry describes the non-trivial topology of these materials ~\cite{Teo}. Additionally, SnS, SeTe, $Ca_2As$ family, and antiperovskites have also been reported to be TCIs ~\cite{Tanaka, Dziawa, Xu2, Zhou, Hsieh2}. Also, PbSe, PbTe, and PbS are predicted to be TCIs with a suitable combination of applied pressure or strain ~\cite{Barone}.

The surface states of TCIs against the magnetic field strength seem to be more powerful than that of the TIs due to the mirror symmetry preservation without preserving the time reversal, but it may be ample to break the mirror symmetry and to be a trivial insulator ~\cite{Fu, Munoz}. Therefore, there are plus and minus points with the crystalline protection of TCIs ~\cite{Hsieh,Tanaka, Dziawa, Xu2, Zhou, Hsieh2}, but it is desirable to find new topological materials~\cite{laxs, tmschoop, tmfeldser, tebs} with pronounced transport properties and a wide range of controllability and functionalities for growing potential applications: quantum computing and spintronic devices. 

In this work, a detailed study of electronic band structures and density of states (DOS) of new ternary $B_2X_2$Zn ($X$=Ir, Rh, Co) compound in the tetragonal crystal structure of symmorphic space group 139 [I4/mmm] is presented. To the best of our knowledge, these compounds have not yet been reported either experimentally, or theoretically. In this paper, we report calculated electronic band structures of $B_2X_2$Zn after completing the full volume optimization process by allowing to rearrange of atomic positions and lattice parameters to minimize the energy. We predict the formation of type I multiple Dirac crossings around X and P points on the Brillouin-zone (BZ) near Fermi energy. With the careful investigation of electronic band structures and DOS of three compounds with and without the spin-orbit coupling (SOC), topological features are identified and discussed in detail. Further, we theoretically demonstrate that $B_2X_2Zn$ has a signature of topological crystalline insulating features with mirror symmetry. We study the surface states on the crystal plane (001) by projecting $X\Gamma$ to surface $\overline M$ $\overline\Gamma$. Additionally, elastic constants and formation energy calculations are performed to verify the mechanical and structural stability of the new compounds. Further, the dynamical stability of compounds is assured by conducting phonon mode calculations.

\section{Computational Method}
The first principle density functional theory (DFT) calculation implemented on Quantum ESPRESSO (QE) simulation package ~\cite{QE1, QE2} has been performed. The plane wave pseudo-potential method formulated with generalized gradient approximation (GGA) and the Perdew-Burke-Ernzerhof (PBE) scheme has been employed with the ultra-soft pseudo-potentials from PSlibrary, including fully relativistic ultra-soft pseudo-potentials for SOC~\cite{Perdew, Singh, Sjostedt, Madsen, Hohen, Kohn}. The k-mesh of 20 x 20 x 20, kinetic energy cutoff for wave functions of 80Ry, and kinetic energy cutoff for charge density and potential of 480Ry are held to receive extreme convergence of energy and charge to enhance the accuracy of the simulation. Additionally, WEIN2K simulation package ~\cite{Blaha, Blaha1} with PBE pseudo-potentials and plane-wave basis set with GGA is used to compare and verify the QE results. Phonopy software package interfaced with QE under harmonic approximation has been used to calculate the phonon spectrum~\cite{Phonon} and ElaStic software package cooperated with QE has been used to calculate the full second-order elastic stiffness tensor~\cite{Elastic}. The crystal momentum in units of $k=(\pi/a, \pi/a, \pi/c)$ is used throughout the discussion unless otherwise specified.   

\section{Results and Discussion}
\subsection{Crystal Structure}
Electron configurations of B, Zn, Ir, Rh, and Co are [He]2$s^2$ 2$p^1$, [Ar]3$d^10$ 4$s^2$, [Xe]4$f^14$ 5$d^7$ 6$s^2$, [Kr] 4$d^8$ 5$s^1$ and, [Ar]3$d^7$ 4$s^2$ respectively. The B-$p$ orbitals, X(Ir, Rh, Co)-$d$ orbitals, and Rh-$s$ orbitals are not fully occupied. The $B_2X_2$Zn ($X$=Ir, Rh, Co) compounds are structured theoretically as a tetragonal crystal structure of space group 139 [I4/mmm], which is belong to the centrosymmetric symmorphic space group. 

\begin{figure}[!htbp]
  \centerline{\includegraphics[width=0.50\textwidth]{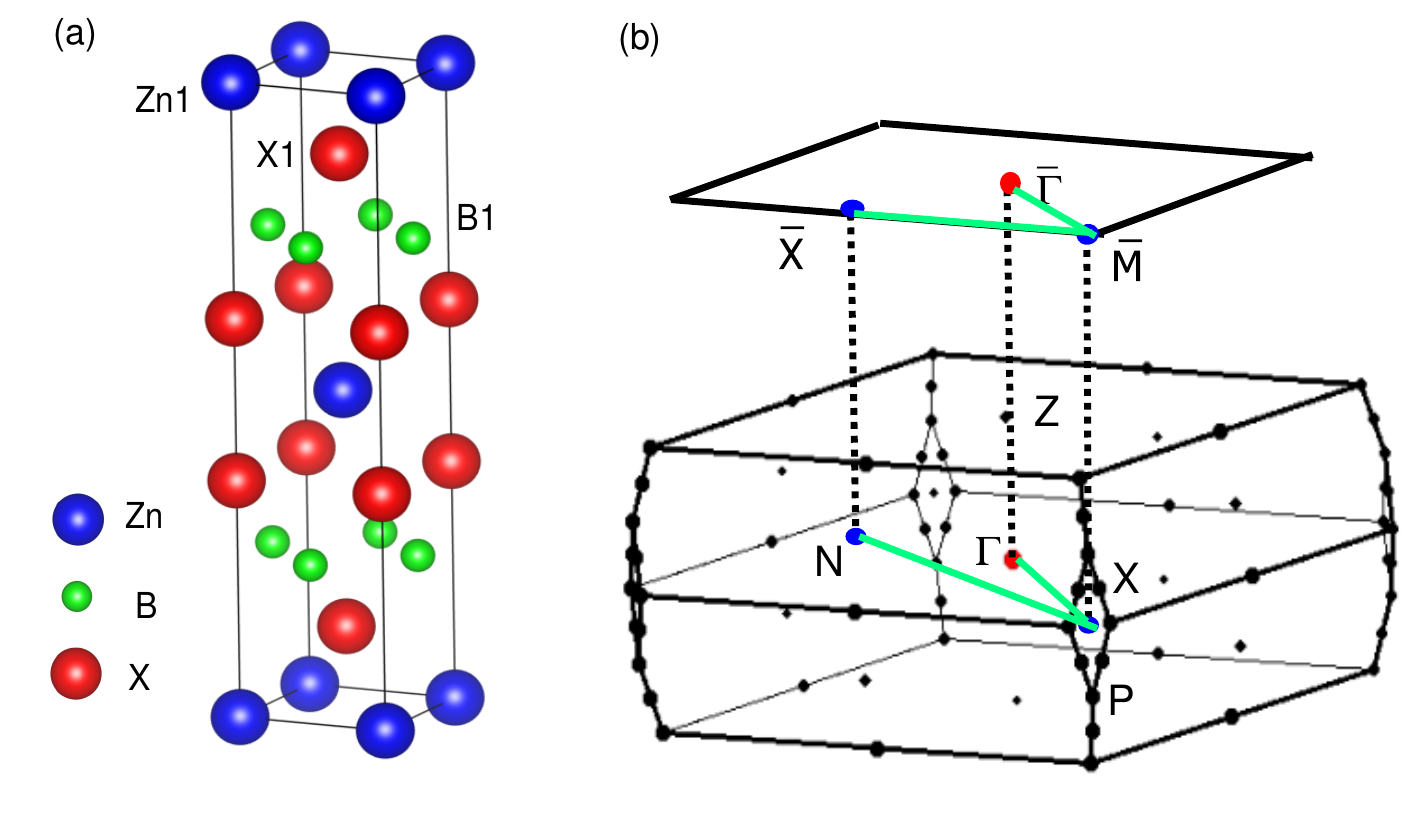}}
  \caption
    {(Color online) Crystal structure and BZ of B$_2X_2$Zn compound. (a) shows a tetragonal crystal structure of a layered pattern. The blue, green, and red solid spheres denote the Zn, B, and X atoms, respectively. The arrangement of atoms for each layer (Zn1, X1, and B1) is noted as top–down views. (b) shows the first BZ by displaying high symmetry points of the BZ (black dots) with labeling $\Gamma$(0, 0, 0), X(-0.5, -0.5, 0), P(0.75, -0.25, -0.25), and Z(0.5, 0.5, -0.5). The (001) surface BZ shows in labeling with $\overline\Gamma$, $\overline M$, and $\overline X$. Paths labeled in green display the k-path selection for the comparison of bulk and surface states calculations. 
    }
\label{Structure}
\end{figure}

The B$_2X_2$Zn structure is shown in Fig.~\ref{Structure} by denoting Zn, B, and X in blue, green, and red colored spheres respectively. The arrangement of atoms for Zn1, X1, and B1 layers is labeled as top–down views. The first BZ of the structure shows in Fig.~\ref{Structure} (b) with the high symmetric points and  $\Gamma$(0, 0, 0) point located at the center. The (001) surface BZ shows in labeling with $\overline\Gamma$, $\overline M$, and $\overline X$ including surface BZ $\overline\Gamma$ at the center.

\subsection{Volume Optimization}
Since the volume that shows the lowest total energy can be identified as the ground state for the stable structure, first we optimize the crystals to find the cell volume that has the lowest total energy. In our calculations, we performed the geometry optimizations of the unit cell by allowing atomic positions to rearrange and the lattice parameters to move around together. We follow the Quantum Espresso volume optimization scheme with extreme convergence for non-relativistic ultra-soft pseudo-potentials. The optimized  lattice constants a and c for B$_2X_2$Zn are shown in the second and third columns of Table.~\ref{tab: Formation}. Similarly calculated lattice positions for the optimized system is Zn(0, 0, 0), B1(0.25, 0.75, 0.50), B2(0.75, 0.25, 0.50), with a little differentiation of the Ir, Rh, and Co positions as Ir1(x, x, 0), Ir2(1-x, 1-x, 0), Rh1(y, y, 0), Rh2(1-y, 1-y, 0), Co1(z, z, 0), and Co2(1-z, 1-z, 0) with x=0.3725808923, y=0.3715044822, and z=0.3712045162. Hence, the crystal structures at optimized bulk lattice parameters and atomic positions are used for further investigation of B$_2X_2$Zn properties during the project. 

\subsection{Formation Energy}
Since the compounds that show negative formation energies at thermal equilibrium with respect to their elemental phases are known to be stable, we calculate the formation energy to study the stability of the structures. In general, the formation energy per atom for ternary B$_2X_2$Zn can be calculated as 
\begin{eqnarray}
  \label{Eq:FE} \nonumber E_f^{B_2X_2Zn} &=& \frac{E^{B_2X_2Zn}-N_{B}E^{B}-N_XE^{X}-N_{Zn}E^{Zn}}{N_{B}+N_X+N_{Zn}}\\
  &=& \frac{E^{B_2X_2Zn}-2E^{B}-2E^{X}-E^{Zn}}{5},
\end{eqnarray}
where $N_B$, $N_X$, and $N_{Zn}$ are the numbers of B, $X$ (Ir, Rh, Co), and Zn atoms in the unit cell, respectively. Since the B$_2X_2$Zn unit cell has 1 atom of Zn and 2 atoms of B and X, $N_{Zn}=1$ and $N_X=N_B=2$ are taken. $E^{B_2X_2Zn}$ is the calculated total free energy of the compound, and $E^B$, $E^X$, and $E^{Zn}$ are the calculated total free energies per atom of the elemental phases of B, $X$, and Zn, respectively.

\begin{table}[!htb]
\caption{\label{tab: Formation}The calculated lattice constants a, c, and formation energies E of B$_2X_2$Zn compounds}
\begin{tabular*}{\columnwidth}{@{\extracolsep{\fill}}rccr}
\toprule
  & a(\AA)& c(\AA)& E(eV/atom)\\
\hline
$B_2Ir_2Zn$$\:$& 2.9912   & $\:$12.7118   & -0.3005\\
$B_2Rh_2Zn$& 2.9377   & $\:$12.9377     & -0.3245\\
$B_2Co_2Zn$& 2.7634  & 12.1513     & -0.2279\\
\botrule
\end{tabular*}
\end{table}

During total energy calculation of B, $X$, and Zn, we use optimized structures of trigonal crystal of space group 166 (R$\tilde{3}$m) for B, a cubic crystal of space group 225 (Fm$\tilde{3}$m) for $X$, and hexagonal crystal of space group 194 ($P6_3$/mmc) for Zn. By using~ Eq. ~\ref{Eq:FE} we calculate the formation energy for $B_2X_2Zn$ as -0.3005 eV/atom, -0.3245 eV/atom, and -0.2279 eV/atom, for $X$=Ir, Rh, and Co respectively, the last column of Table.~\ref{tab: Formation}. Since all three compounds indicate negative formation energies with respect to their elemental phases, we identify those structures are theoretically stable.

\subsection{Elastic Properties}
First principles density functional calculation implemented on QE is used to explore the mechanical properties of the structures. Mechanical properties together with crystal stability and stiffness are easily investigated by calculating the elastic stiffness matrix $C_{ij}$ or flexibility matrix $S_{ij}$ ($=[C_{ij}]^{-1})$. The bulk modulus, Young’s modulus, shear modulus, and Poisson’s ratio of polycrystals are calculated by Voigt-Reuss approximation methods~\cite{Elastic}. Average poly-crystalline modules (Hill's average) are obtained by using the upper and lower limit of the actual effective modulus corresponding to the Voigt bound and Reuss bound. The Hill's average is said to be mostly agreed with the experimental result~\cite{Elastic}.
Since the structure is a tetragonal crystal, there are six independent non-zero elastic constants, namely $C_{11}, C_{12}, C_{13}, C_{33}, C_{44},$ and $C_{66}$. The calculated elastic constants and effective bulk, shear, and Young's modules for $B_2X_2Zn$ are presented in Table.~\ref{tab: Elastic}. 

The elastic constants calculated for the tetragonal crystal satisfy the following Born mechanical stability criteria as discussed in~\cite{Born1, Born2}:
\begin{eqnarray}
  \label{Eq:Born} 
  \left\{ 
  \begin{array}{ll}
 C_{11}>C_{12};& 2C_{13}^2<C_{33}(C_{11}+C_{12})\\
  C_{44}>0;& C_{66}>0,
  \end{array}
 \right. 
\end{eqnarray}
where $C_{ij}$ represent six independent non-zero elastic constants. The elastic tensor of the second order is calculated by using the expansion of the elastic energy in terms of the applied strain. The $C_{ij}$ results are obtained for large deformations with high-order polynomial fit by identifying the plateau regions which provides good reasonable results~\cite{Elastic}. Calculated elastic constants for all three $B_2X_2Zn$ compounds satisfy the Born mechanical stability criteria implemented in Eq. ~\ref{Eq:Born}.

\begin{table*}[!htb]
\caption{\label{tab: Elastic} The calculated elastic constants ($C_{ij}$), bulk modulus (B), shear modulus (G), Young’s modulus (E) in units of GPa, and Poisson’s ratio ($\nu$) for the B$_2X_2$Zn compounds.}
\begin{tabular*}{\textwidth}{@{\extracolsep{\fill}}rcccccccccr}
\toprule
  & $C_{11}$&$C_{12}$&$C_{13}$&$C_{33}$&$C_{44}$&$C_{66}$&$B$&$G$&$E$&$\nu$\\
\hline
$B_2Ir_2Zn$$\:$& 354.4   & 142.2    &199.8   & 496.5  & 183.3 &100.0    & 247.92   & 132.07& 336.47  &0.27\\
$B_2Rh_2Zn$& 338.4   & 102.6   &169.2   & 420.1  & 157.9 &80.4    & 215.81   & 118.47& 300.43  &0.27\\
$B_2Co_2Zn$& 342.1   & 133.8    &160.7   & 432.3  & 205.8 &120.3    & 223.03   & 143.92& 355.32 &0.23\\
\botrule
\end{tabular*}
\end{table*}

\subsection{Phonon Frequencies}
A collective excitation of a set of atoms that decomposed into different modes plays an essential role in material science. Therefore, we perform the first principles of phonon calculations with force constants which are said to be an important calculation for studying dynamical behaviors and thermal properties of the materials. Studying topological phonons and their properties is another frontier research field that we are not discussing here. Our focus here is to check the dynamical stability of the samples by observing non-imaginary phonons frequencies. The vibrational band structures of B$_2X_2$Zn do not have regions with the imaginary frequency which imply that the phonon stability of the samples. It should be noted that the phonon calculation is the most widely used in the stability analysis, therefore those compounds are likely to be realized experimentally.

\begin{figure}[!htp]
  \centerline{\includegraphics[width=0.5\textwidth]{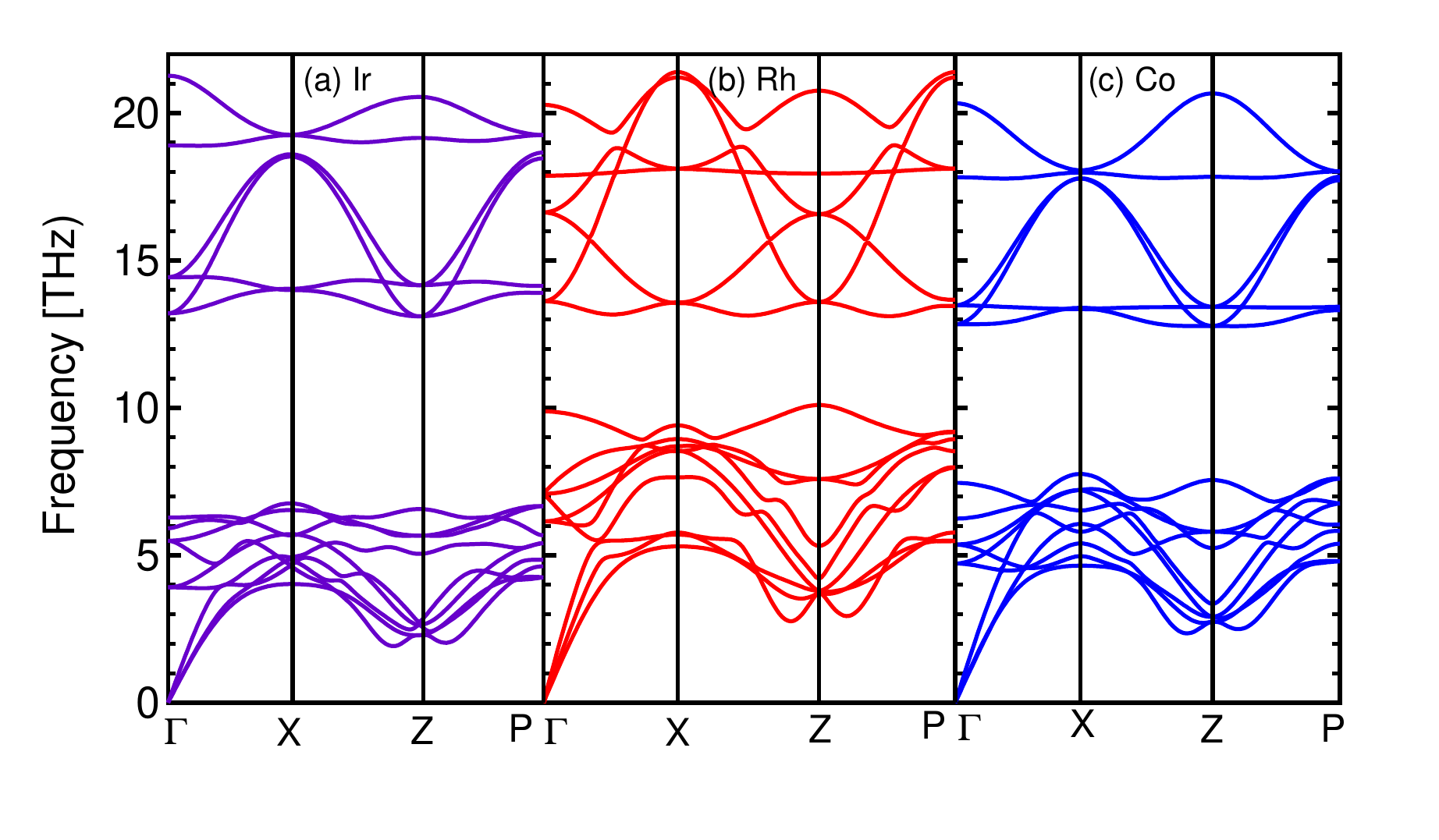}}
  \caption
    {(Color online) Calculated phonon spectra of B$_2X_2$Zn ($X= Ir, Rh, Co$) compounds. All three phonon spectra were calculated without SOC effect and are shown on k-path $\Gamma-X-Z-P$. (a) represents the phonon spectrum of the Ir sample, (b) represents the same for the Rh sample, and (c) represents the same for the Co sample. 
   }
\label{Phonon}
\end{figure}
\vspace{10cm}
 Due to the negative formation energies with the fulfillment of the mechanical stability scheme and non-imaginary phonon spectra of all three $B_2X_2Zn$ compounds, we conclude that all three compounds are mechanically, structurally, and dynamically stable.

\subsection{Electronic Band Structure and DOS Properties}
\begin{figure}[!htb]
  \centerline{\includegraphics[width=0.5\textwidth]{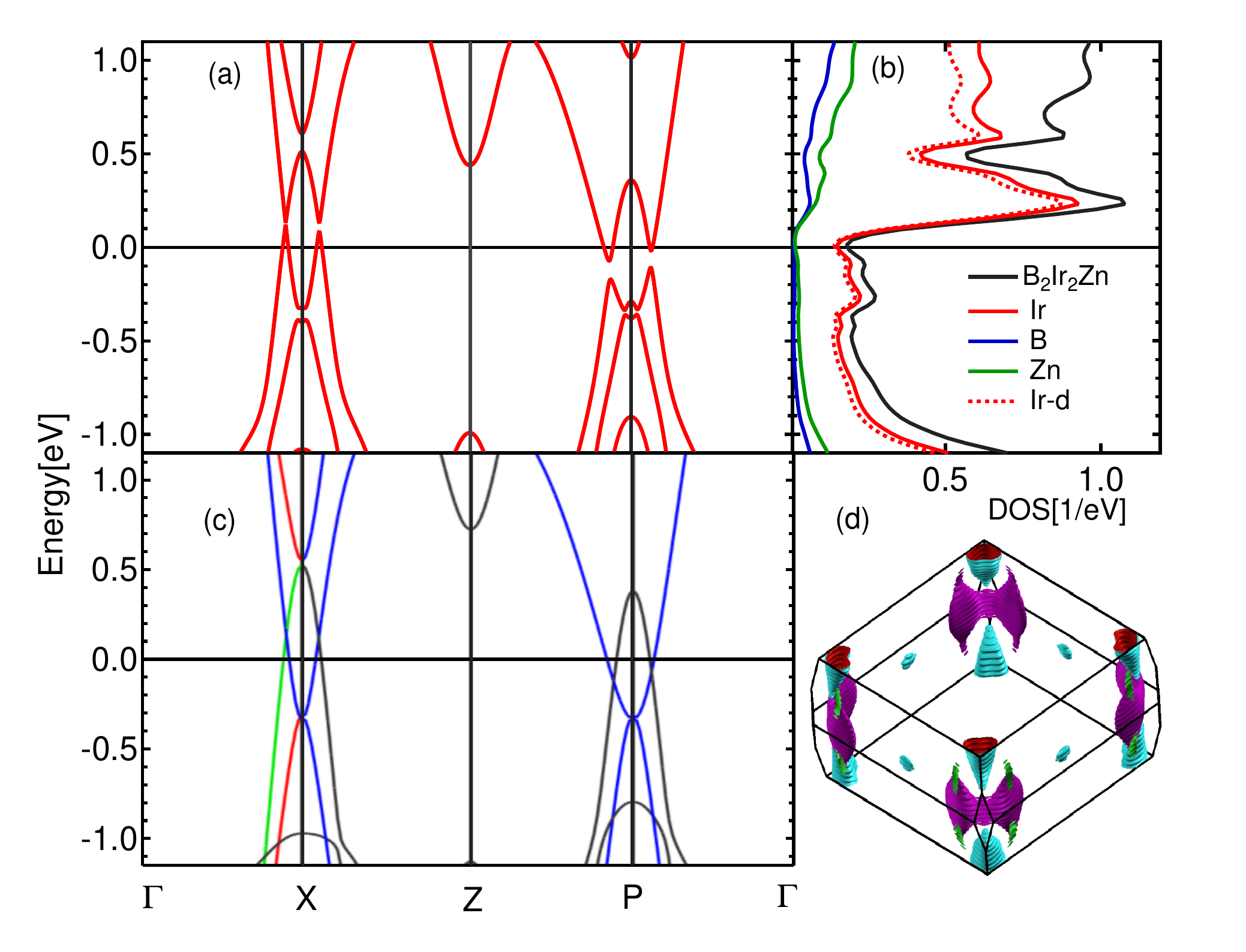}}
  \caption
    {(Color online) Calculated electronic band structure of B$_2Ir_2$Zn compound with and without SOC interaction along the high symmetry lines on the BZ k-path $\Gamma-M-X-Z-P-\Gamma$. (a) shows Calculated bulk band structure with SOC effect. (b) Calculated atom-projected DOS for B$_2Ir_2$Zn with SOC and total DOS with SOC effect. (c) represents an irreducible representation of the Calculated bulk band structure without SOC. Colors have the meaning of band symmetries discussed in the text. The solid black line at zero in (a), (b), and (c) indicates the Fermi level. (d) shows the Fermi surface of B$_2Ir_2$Zn bands.
   }
\label{BIrZn}
\end{figure}

Electronic band structure calculation of B$_2Ir_2$Zn within GGA with and without the inclusion of the SOC along high-symmetry k-path $\Gamma-X-Z-P-\Gamma$ are plotted by setting the Fermi level at 0 eV on energy scale as shown in Fig.~\ref{BIrZn}. The top panel (a) shows the band structure of B$_2Ir_2$Zn compound with the inclusion of the SOC effect and the bottom panel (c) shows the calculated electronic band structure without the inclusion of the SOC effect in irreducible representation. Irreducible representation (symmorphic crystal symmetries) of band structure shows band symmetries by using different colored solid lines. There are few bands near the Fermi level. Interesting band features near the Fermi level are noted around X and P points. Since irreducible representation allows us to access each eigenvalue along the chosen k-path, we can identify connecting lines of bands and the symmetries by looking for  the same colored bands for the same symmetry~\cite{Kittle, Koster}.  

The two linear upright crossings in the $\Gamma$-X-Z plane are identified as Dirac-like crossings. Since the SOC (some of the degenerate atomic levels split without magnetic field) have appeared as promising candidates for exotic band behaviors of Dirac materials, we perform SOC calculation to identify the topological features at the crossings. As shown in the top panel (a) in Fig.~\ref{BIrZn}, it is clear that those crossings are gapped out with the inclusion of SOC. The Dirac point located at (0.136 eV) with the coordination of ~$k$=(0.499, -0.499, 0) is gaped out into two-fold degeneracy. Blue color represents the $\Gamma_2$ and Green color represents the $\Gamma_4$ in irreducible representations with space group $C_{2v}$. The Dirac point located at (0.110 eV) is gaped out into two-fold degeneracy. Blue color represents the $\Gamma_2$ and black color represents the $\Gamma_1$ in irreducible representations with space group $c_s$. Therefore, both crossings are identified as type I two-fold degenerate Dirac points. The lines at crossing points display the same and opposite slopes around $\pm$7 eV \r{A}. These Dirac points are protected by the absence of SOC with the predicted electron velocity of around $1.0\times10^{16}$ \r{A} /s calculated by $(1/\hbar) dE/dk$, which is similar to the experimentally measured velocity of Cd$_3$As$_2$~\cite{Liang}.

The two crossings located at the Z-P-$\Gamma$ plane (left at -0.128eV and right at -0.061eV) are gapped out again into two-fold degeneracy with band inversion. At the left crossing, black and blue bands represent the $\Gamma_1$ and $\Gamma_2$ in an orderly space group $C_s$. At the right crossing, black and blue colors represent the $\Gamma_1$ and $\Gamma_2$, respectively with space group $C_s$. Therefore, both crossings at around -0.128eV  and -0.061ev are identified as type I two-fold degenerate Dirac points.  The lines at crossing points display the same and opposite slopes around $\pm$5 eV \r{A}. These Dirac points are protected by the absence of SOC but gap out by the presence of SOC. It is also observed that there is a band inversion during the gap out. All the band crossing near Fermi energy are identified mainly as Ir$-d$ orbitals with barely hybridized with all$-p$ orbitals by using the fat band orientations.

The results of total and partial DOS of B$_2Ir_2$Zn provide valuable information about the origin of bands with contributions from each atom and each orbital. Total and atom-projected DOS calculations without SOC effect for B$_2Ir_2$Zn display in Fig.~\ref{BIrZn}(c). Black solid lines represent the total DOS from all atoms of the B$_2Ir_2$Zn compound. Red, blue, and green represent the atom-projected total DOS for Ir, B, and Zn atoms respectively. The total DOS at the Fermi level is around 0.25 states per eV per unit cell and is dominated by Ir atom DOS. Further, total DOS at the Fermi level is dominated by Ir$-d$ orbitals denoted by red dotted lines. This is agreed with the fat band orientations we discussed in the bands' diagram in Fig.~\ref{BIrZn}.  

The calculated Fermi surfaces display in Fig.~\ref{BIrZn} (d) by indicating cone-shaped Fermi pockets at the corner of the Fermi surface within the BZ.  

\begin{figure}[!htb]
  \centerline{\includegraphics[width=0.5\textwidth]{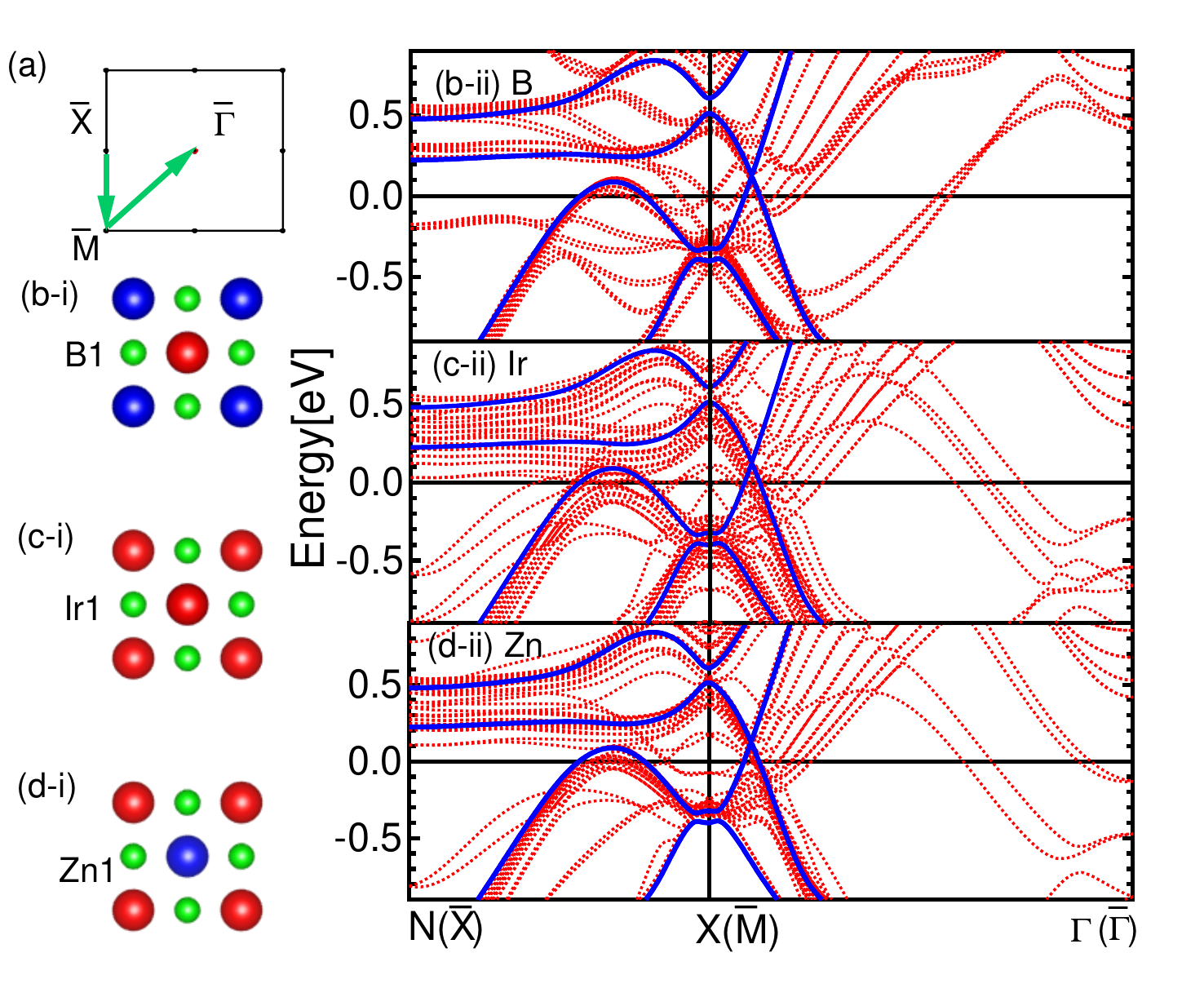}}
  \caption
     {(Color online) Calculated surface states of (001) plane. (a) The (001) surface BZ shows in labeling with $\overline\Gamma$, $\overline M$, and $\overline X$. Paths labeled in green display the k-path selection.
     (b-i) shows the arrangement of atoms for the B1 layer and (b-ii) shows the surface states of the B-termination layer with 25 layers of atoms. (c-i) shows the arrangement of atoms for the Ir1 layer and (c-ii) shows the surface states of the Ir-termination layer with 21 layers of atoms. Similarly, (d-i) shows the arrangement of atoms for the Zn1 layer, and (d-ii) shows the surface states of the B-termination layer with 23 layers of atoms. The arrangement of atoms for each layer (Zn1, X1, and B1) is noted in Fig.~\ref{Structure}. Surface bands are denoted by red solid lines and bulk bands are shown in blue solid lines as a comparison. All surface and bulk band calculations are performed including the SOC effect. The solid black horizontal line at zero indicates the Fermi level.
   }
\label{Surface}
\end{figure}
This work is extended to the classification of band structures in a different direction including crystal point group symmetries. Recent studies of topological materials reveal a large class of materials with gapless surface states. Among them, topological insulating states protected by crystal symmetries, rather than time-reversal symmetry are introduced as topological crystalline insulators. This motivates us to investigate topological crystalline insulators' behaviors of the compound. As we show in Fig.~\ref{Structure} (c), the projection of N-X-$\Gamma$ in bulk BZ represents from $\overline X$-$\overline M$- $\overline\Gamma$ in (001) surface BZ. In Fig.~\ref{Structure}, (a) again represents the (001) surface BZ. Calculated surface states on k-Path labeled in green display on the right top to bottom by using red solid lines for different termination layers as in (b-i), (c-i), and (d-i) of Fig.~\ref{Surface}. To study the surface states compared to bulk bands, we plot the bulk band on top of the surface bands by using blue solid lines. All the calculations were performed for relativistic potentials. We found that all B, Ir, and Zn termination layers show the same features of surface states although they show different band densities. Additionally, we investigate clear surface states on the X-N plane since there is no crossing observed in bulk bands. The surface states on the X-$\Gamma$ plane are barely recognized due to small gap openings of bulk bands. As you can see we are unable to project the interested exact bulk band k-path in Fig.~\ref{BIrZn} to any surface directly. That was the reason to look at the (001) surface plane with the projection of N-X-$\Gamma$.

\begin{figure}[!htbp]
  \centerline{\includegraphics[width=0.55\textwidth]{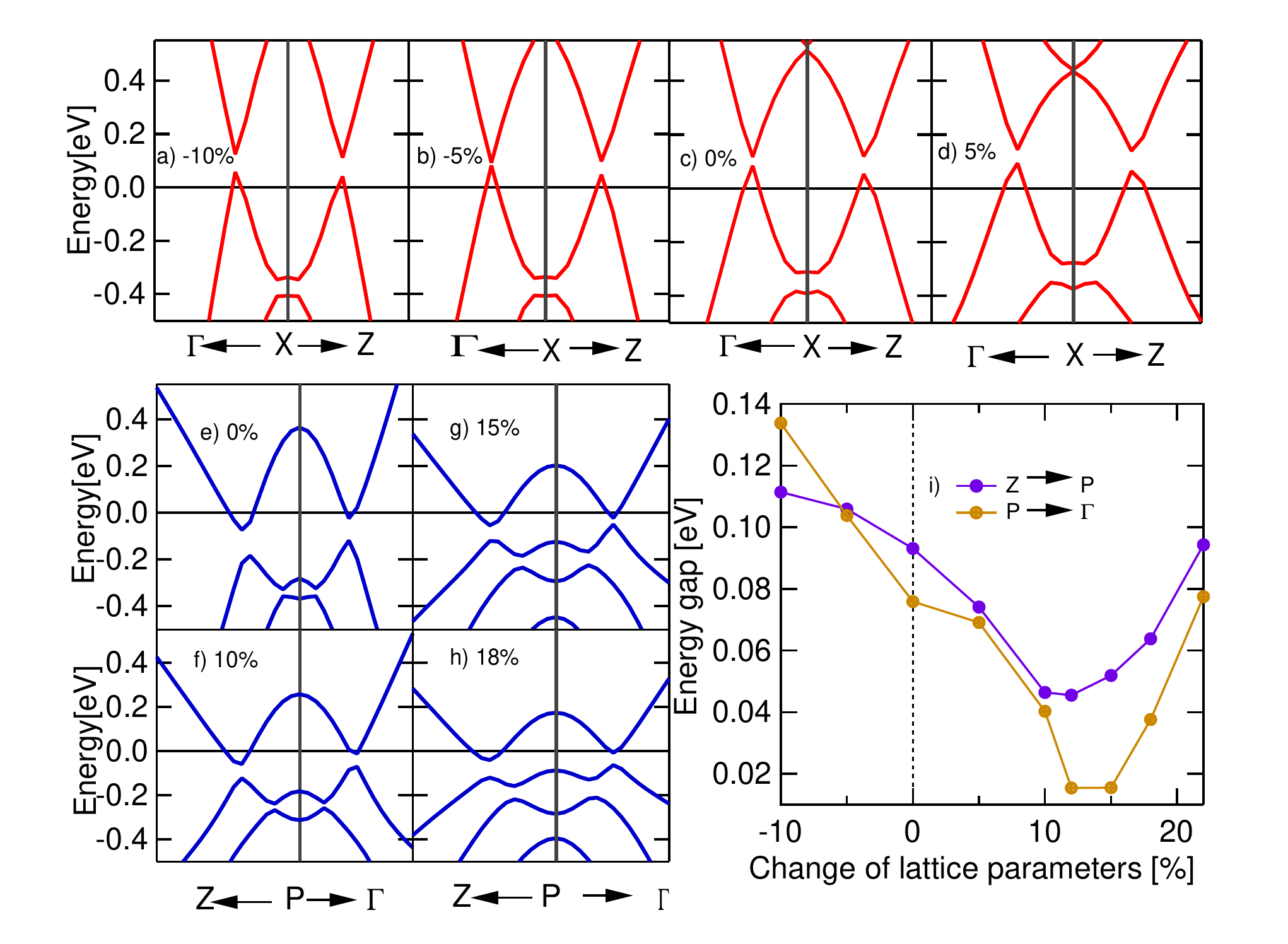}}
 \caption
    {(Color online) the change in band energy gap around the Fermi level of the B$_2Ir_2$Zn sample with SOC effect by changing the volume of the lattice. Top panel: shows the effect of gap changes due to the compression and stretch of the lattice for two crossings in X-$\Gamma$ and X-Z paths for -10\%, -5\%, 0\%, and 5\% showing left to right respectively. Bottom panel: The plots on the left show two crossings in the Z-P-$\Gamma$ k-path with their related stretch percentages in volume. The graph on the right compares both crossings' energy gap to the percentage of lattice parameters in P-Z and P-$\Gamma$ paths on the left. The dotted back vertical line at 0\% represents the optimized lattice.  The solid black horizontal line at zero indicates the Fermi level in all plots. 
   }
\label{Pressure}
\end{figure}
We further studied the B$_2Ir_2$Zn by investigating the dependence of the band gap on the lattice constants. We investigate the band gap as a function of the percentage of lattice parameters. The behavior of the two crossings identified as Dirac points on $\Gamma$-X-Z plane in Fig~\ref{BIrZn} is investigated as shown in the top panel of  Fig~\ref{Pressure}. It displays that there is no manifest effect to change of lattice parameters even by compressing or stretching. We were barely able to identify the gap closing of the X-$\Gamma$ plane around 5\% of lattice compressing, but no effect was identified on the X-Z plane. The behavior of the two crossings identified as Dirac points with band inversion Z-P-$\Gamma$ path in Fig~\ref{BIrZn} is investigated as shown in the bottom left panel of Fig~\ref{Pressure}. It displays that there is a considerable effect on band gap with respect to lattice parameters by stretching. As the lattice parameters increase, the band gap of B$_2Ir_2$Zn decreases to zero and then re-opens. This gap closing signals a topological crystalline insulator at that ambient pressure which is expected to close 13\% of stretching the lattice. The energy gap as a function of the percentage change of lattice parameters is displayed in Fig~\ref{Pressure} (i) by denoting purple solid color for Z-P and brown solid color for P-$\Gamma$ paths. Continuously tunable band gaps may lead to wide-ranging applications in
thermoelectrics, infrared detection, and tunable electronics.

\begin{figure}[!htp]
  \centerline{\includegraphics[width=0.5\textwidth]{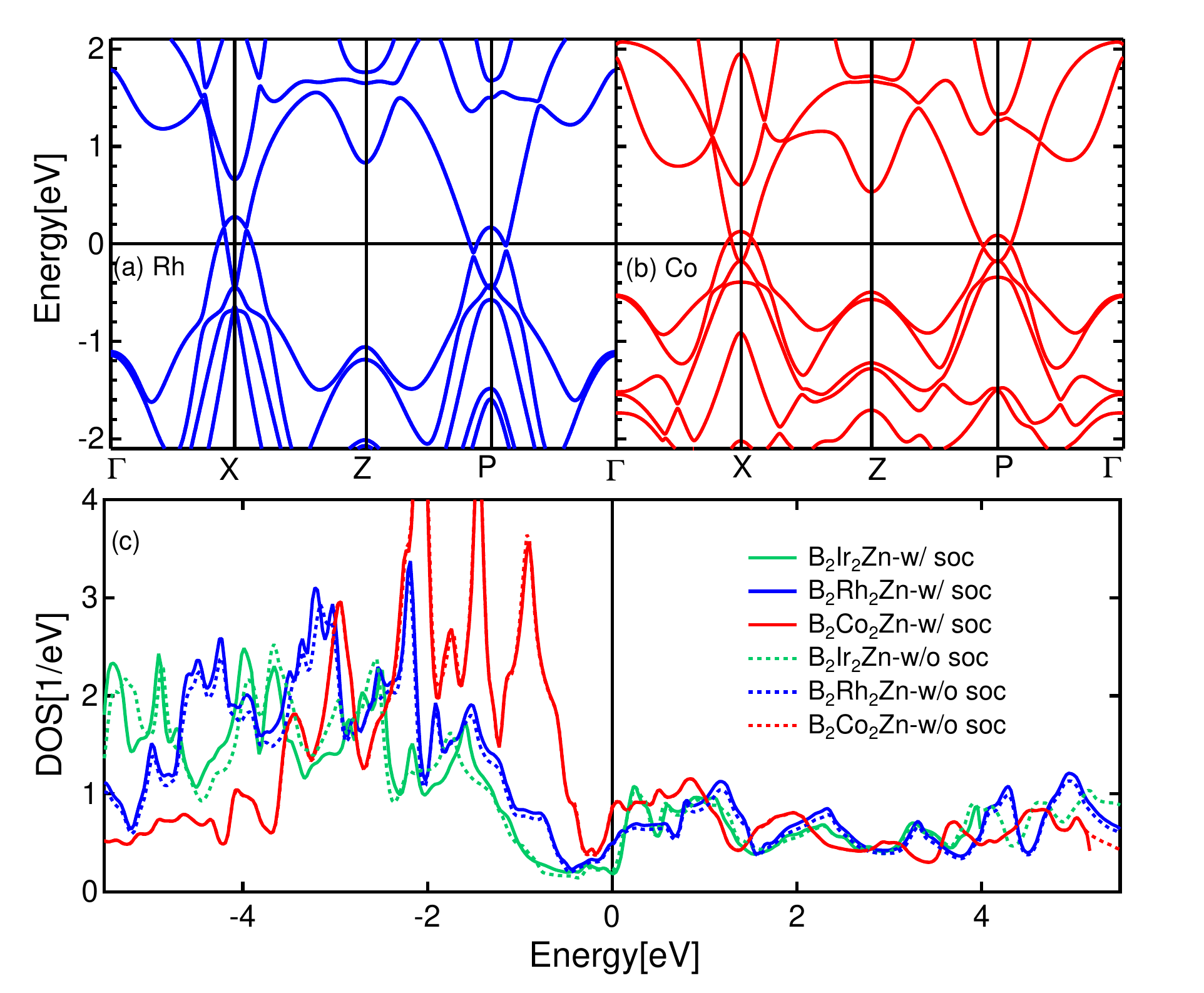}}
  \caption
    {(Color online) Calculated band structure of B$_2X_2$Zn ($X=Rh, Co$) without SOC, and DOS of B$_2X_2$Zn ($X=Ir, Rh, Co$) with and without SOC effect. (a) and (b) display the band structure of Rh and Co samples respectively. (c) shows total DOS for all three samples representing solid lines for with and dotted lines for without SOC effect. The solid black line at zero indicates the Fermi level in all plots.
   }
\label{DOS}
\end{figure}

 We perform the electronic band structure and DOS calculations for B$_2Rh_2$Zn and B$_2Co_2$Zn by using the same structure as B$_2Ir_2$Zn. All the calculations are done by using the optimized lattice parameters from Table .~\ref{tab: Formation} and including the SOC effect as shown in (a) and (b) of Fig.~\ref{DOS}. We conclude that all three compounds display the same band characteristics as discussed above for B$_2Ir_2$Zn. The energy locations of four crossing points are different than the B$_2Ir_2$Zn. It is also identified that band structure is more compressed toward the Fermi energy in Co and Rh than Ir.    
 The DOS features around the Fermi level are investigated in three samples with and without SOC effect as shown in (c) of Fig.~\ref{DOS}. We did not identify the manifest SOC effect in DOS by comparing solid and dotted color lines represented with and without relativistic potentials. Fig.~\ref{DOS} shows that Co has more DOS near Fermi energy than Rh and Rh has more DOS than Ir. 
\hspace{10cm}
\section{Conclusion}
In summary, newly designed B$_2X_2$Zn ($X= Ir, Rh, Co$) compounds show interesting topological properties and the hallmark of the topological crystalline insulator. Dirac band behaviors near the Fermi level with the effect of  spin-orbit coupling can be used to predict theoretical and experimental significance on material properties. Linear band crossings near the Fermi level are discovered. Type I Dirac crossings are identified by investigating the SOC effect. We perform the surface states calculation on high symmetric crystal surfaces by projecting to the (001) plane. The gaping behavior is quantitatively investigated by applying stress and strain to the lattice volume. It shows that gap is closing by stretching the lattice and also re-opening for further stretching which is identified as the topological crystalline insulating behavior. Further, calculated elastic constants, formation energy, and phonon spectra predict the mechanical, structural, and dynamical stabilities of the compounds. The predicted electronic structures of B$_2X_2$Zn compounds, their important topological properties, and stability criteria will be useful for investigating further studies. Since these materials show high mobilities and a wide range of functionalities, it creates an extremely valuable platform for exploring advanced technology.

\section{Acknowledgment}
NH and KH acknowledge the Extreme Science and Engineering Discovery Environment (XSEDE), supported by grant number TG-PHY190050. KH and JH acknowledge the financial support from Undergraduate Prestigious Fellowships from Seton Hall University.


\begin{thebibliography}{999}
\bibitem{Kane} C. L. Kane and E.J. Mele,  $Z_2$ topological order and the quantum spin hall effect. {\em Phys. Rev. Lett.} {\bf 2005}, {\em 95}, 146802.
\bibitem{Kane2} C. L. Kane and E.J. Mele, Quantum spin hall effect in graphene. {\em Phys. Rev. Lett.} {\bf 2005}, {\em 95}, 226801.
\bibitem{Hasan}M. Z. Hasan, {\it et al.}, Topological insulators. {\em Rev. Mod. Phys.} {\bf 2010},{\em 82}, 3045.
\bibitem{Qi}X. L. Qi, {\it et al.}, Topological insulators and superconductors. {\em Rev. Mod. Phys.} {\bf 2011}, {\em 83}, 1057.
\bibitem{Yang1}B.-J. Yang, {\it et al.}, Classification of stable three-dimensional Dirac semimetals with nontrivial topology. {\em Nat. Commun.} {\bf 2014} {\em 5}, 4898.
\bibitem{Xu1}G. Xu, {\it et al.}, Chern Semimetal and the Quantized Anomalous Hall Effect in $HgCr_2Se_4$. {\em Phys. Rev. Lett.} {\bf 2011}, {\bf 107}, 186806.
\bibitem{Zyuzin}A. A. Zyuzin, {\it et al.}, Weyl semimetal with broken time reversal and inversion symmetries. {\em Phys. Rev. B.} {\bf 2012}, {\em 85}, 165110.
\bibitem{Mong} R. S. K. Mong, {\it et al.}, Antiferromagnetic topological insulators. {\em Phys. Rev. B.}{\bf 2010}, {\em 81}, 245209.
\bibitem{Teo} J. Y. C. Teo, {\it et al.}, Surface states and topological invariants in three-dimensional topological insulators: {\em Phys. Rev. B.} {\bf 2008}, {\em 78}, 045426.
\bibitem{Xia} Y. Xia, {\it et al.} Observation of a large-gap topological-insulator class with a single Dirac cone on the surface. {\em Nat. Phys.} {bf 2009}, {\em 5} 398.


\bibitem{Chen} Y. L. Chen, {\it et al.} Experimental Realization of a Three-Dimensional Topological Insulator, $Bi_2Te_3$. {\em Science}, {\bf 2009} {\em 325} 178.
\bibitem{Hsieh1} D. Hsieh, {\it et al.} Observation of Time-Reversal-Protected Single-Dirac-Cone Topological-Insulator States in $Bi_2Te_3$ and $Sb_2Te_3$. {\em Phys, Rev. Lett.}{\bf 2009}, {\em 103}, 146401.
\bibitem{Hsieh} T. H. Hsieh, {\it et al.} Topological crystalline insulators in the SnTe material class. {\em Nat. Commun.} {\bf 2012}, {\em 3}, 982.
\bibitem{Tanaka} Y. Tanaka, {\it et al.} Experimental realization of a topological crystalline insulator in SnTe. {\em Nat. Phys.} {\bf 2012}, {\em 8}, 800.
\bibitem{Fu} L. Fu, Topological crystalline insulators. {\em Phys. Rev. Lett.}
{\bf 2011}, {\em 106}, 106802.
\bibitem{Fang} C. Fang and L.Fu, New classes of topological crystalline insulators having surface rotation anomaly.{\em Science Advances.} {\bf 2019}, {\em 5}.
\bibitem{Weng} H. Weng, {\it et al.} Topological crystalline kondo insulator in mixed valence ytterbium borides. {\em Phys, Rev. Lett.}{\bf 2014}, {\em 112}, 016403.
\bibitem{Dziawa} P. Dziawa. {\it et al.} Topological crystalline insulator states in $Pb_{1-x}Sn_xSe$. {\em Nat. Mat.} {\bf 2012}, {\em 11}, 1023.
\bibitem{Xu2} S. Xu, {\it et al.} Observation of a topological crystalline insulator phase and topological phase transition in $Pb_{1-x}Sn_xTe$. {\em Nat. Commun.} {\bf 2012}, {\em 3}, 1192.
\bibitem{Zhou} X. Zhou, {\it et al.} Topological crystalline insulator states in the $Ca_2As$ family, {\em Phys. Rev. B.} {\bf 2018}, {\em 98}, 241104.

\bibitem{Hsieh2} H. Hsieh, {\it et al.} Topological crystalline insulators and Dirac octets in antiperovskites {\em Phys. Rev. B.} {\bf 2014}, {\em 90}, 081112.
\bibitem{Barone} P. Barone, {\it et al.}, Pressure-induced topological phase transitions in rocksalt chalcogenides. {\em Phys. Rev. B.} {\bf 2013}, {\em 88}, 045207.
\bibitem{Munoz}F. Munoz, {\it et al.}, Topological Crystalline Insulator in a New Bi-Semiconducting Phase, {\em Scientific Rep.} {\bf 2016}, {\em 6}, 21790.
\bibitem{laxs}J. Howard, {\it et al.}, Computational Prediction of New Series of Topological Ternary Compounds LaXS (X = Si, Ge, Sn) from First-Principles. {\em J} {\bf 2021}, {\em 4}, 577-588, 
\bibitem{tmschoop} L. M. Schoop, {\it et al.} Topological Materials and Solid-State Chemistry—Finding and Characterizing New Topological Materials {\em Topological Matter. Springer Series in Solid-State Sciences} {\bf 2018}, {\em 190}, 211-243 
\bibitem{tmfeldser} Y. Sun, {\it et al.}Topological Materials in Heusler Compounds {\em Topological Matter Springer Series in Solid-State Sciences} {\bf 2018}, {\em 190}, 199-210
\bibitem{tebs} P. Narang, {\it et al.} The topology of electronic band structures. {\em Nat. Mater.} {\bf 2021} {\em 20}, 293–300 


\bibitem{QE1}P. Giannozzi, {\it et al.}, QUANTUM ESPRESSO: a modular and open-source software project for quantum simulations of materials. {\em J. Phys.: Condens. Matter.} {\bf 2009}, {\em 21}, 395502.
\bibitem{QE2}P. Giannozzi, {\it et al.}, Advanced capabilities for materials modeling with Quantum ESPRESSO. {\em J. Phys.: Condens.Matter.} {\bf 2017}, {\em 29}, 465901.
\bibitem{Perdew} J. P. Perdew, {\it et al.}, Generalized Gradient Approximation Made Simple. {\em Phys. Rev. Lett.} {\bf 1996}, {\em 77}, 3865.
\bibitem{Singh} D. J. Singh; L. Nordstrom, \textit{Planewaves, Pseudopotentials, and the LAPW Method}, 2nd ed.; Springer: USA, 2006; pp. 1--134.
\bibitem{Sjostedt}E. Sjostedt, {\it et al.}, An alternative way of linearizing the augmented plane-wave method. {\em Solid State Commun.} {\bf 2000}, {\em 114}, 15--20. 
\bibitem{Madsen} G. K. H. Madsen, {\it et al.}, Efficient linearization of the augmented plane-wave method. {\em Phys. Rev. B.} {\bf 2001}, {\em 64}, 195134.
\bibitem{Hohen} P. Hohenberg, {\it et al.}, Inhomogeneous Electron Gas. {\em Phys. Rev.} {\bf 1964}, {\em 136}, B864.
\bibitem{Kohn} W. Kohn, {\it et al.}, Self-Consistent Equations Including Exchange and Correlation Effects. {\em Phys. Rev.} {\bf 1965}, {\em 140}, A1133.
\bibitem{Blaha} P. Blaha, {\it et al.}, Full-potential, linearized augmented plane wave programs for crystalline systems. {\em  Commput. Phys. Commun.} {\bf 1990}, {\em 59}, 399--415.
\bibitem{Blaha1} P. Blaha, {\it et al.}, WIEN2k: An APW+lo program for calculating the properties of solids. {\em J. Chem. Phys.} {\bf 2020}, {\em 152}, 074101.
\bibitem{Elastic}R. Golesorkhtabar, {\it et al.}, ElaStic: A tool for calculating second-order elastic constants from first principles. {\em Comp. Phys. Commun.} {\bf 2013}, {\em 184}, 1861-1873.
\bibitem{Phonon}A. Togo, {\it et al.}, First principles phonon calculations in materials science. {\em Scripta Materialia} {\bf 2015}, {\em 108}, 1-5.
\bibitem{Born1} M. Born, On the stability of crystal lattices. I {\em Math. Proc. Camb. Phil. Soc.} {\bf 1940}, {\em 36}, 160--172.
\bibitem{Born2} F. Mouhat, {\it et al.}, Necessary and sufficient elastic stability conditions in various crystal systems. {\em Phys. Rev. B.} {\bf 2014}, {\em 90}, 224104.
\bibitem{Kittle} C. Kittle, \textit{Introduction to solid state Physics}, 8th ed.; John Wiley \& Sons: NY, USA, 2004; pp 1-704.
\bibitem{Koster}G. F. Koster; J. D. Dimmock; R. G. Wheeler; H. Statz, \textit{Properties of the thirty-two point groups}, 1st ed.; MIT Press: Cambridge, MA, 1963; pp 1-104.
\bibitem{Liang} T. Liang, {\it et al.}, Ultrahigh mobility and giant magnetoresistance in the Dirac semimetal Cd$_3$As$_2$. {\em Nat. Mater.} {\bf 2015}, {\em 14}, 280-284.

\end{thebibliography}
\end{document}